\documentclass[letterpaper]{iopart}
\usepackage{graphicx} 
\usepackage{hyperref}
\usepackage{latexsym}

\newcommand{\km}{\hbox{ km}}
\newcommand{\s}{\hbox{ s}}
\newcommand{\mpc}{\hbox{ Mpc}}
\newcommand{\gyr}{\hbox{ Gyr}}

\newcommand{\cfrac}[2]{\textstyle{\frac{#1}{#2}}}
\newcommand{\lcdm}{$\Lambda$CDM}
\def\ltap{\mathop{\raisebox{-.4ex}{\rlap{$\sim$}} \raisebox{.4ex}{$<$}}}
\def\gtap{\mathop{\raisebox{-.4ex}{\rlap{$\sim$}} \raisebox{.4ex}{$>$}}}

\newcommand{\be}{\begin{equation}}
\newcommand{\ee}{\end{equation}}
\newcommand{\bea}{\begin{eqnarray}}
\newcommand{\eea}{\end{eqnarray}}

\catcode`\@=11
\def\ps@fnal{
\def\@oddhead{\textsf{FERMILAB--Pub--05/379-T \hfil \thepage}}
\def\@evenhead{\thepage \hfil \textsf{FERMILAB--Pub--05/379--T}}}
\relax

\begin{document}

 \noindent \textsf{
 \phantom{M} \hfill FERMILAB--PUB--05/379--T}\\[-24pt]

\title{Observational Constraints on  Undulant Cosmologies} 

\author{Gabriela Barenboim}
\address{Departament de F\'{\i}sica Te\`{o}rica,
 Universitat de Val\`{e}ncia, 
Carrer Dr.~Moliner 50, E-46100 Burjassot (Val\`{e}ncia),  ÊÊSpainÊ}
\ead{Gabriela.Barenboim@uv.es}

\author{Olga Mena Requejo}
\address{Theoretical Physics Department, Fermi National 
Accelerator Laboratory,\\ P.O.\ Box 500, Batavia, Illinois 60510 USA}
\ead{omena@fnal.gov}

\author{Chris Quigg}
\address{Theoretical Physics Department, Fermi National 
Accelerator Laboratory,\\ P.O.\ Box 500, Batavia, Illinois 60510 USA}
\ead{quigg@fnal.gov}

\date{\today}

\begin{abstract}
In an undulant universe, cosmic expansion is characterized by 
alternating periods of acceleration and deceleration. We examine 
cosmologies in which the dark-energy equation of state varies 
periodically with the number of $e$-foldings of the scale factor of 
the universe, and use observations to constrain the frequency of 
oscillation. We find a tension between a forceful response to the 
cosmic coincidence problem and the standard treatment of structure 
formation.
\end{abstract}

\submitto{Journal of Cosmology and Astroparticle Physics}




%
\section{Introduction}
Astronomical observations have led to the inference that the Universe
is approximately flat, and that its complement of mass-energy now
consists of 5\% ordinary matter, 22\% nonbaryonic dark matter, and a
dominant negative-pressure component that accelerates the Hubble
expansion~\cite{Peebles:2002gy,Freedman:2003ys,Trodden:2004st}.  The
discovery that the present Universe is expanding at an accelerating pace
arose in measurements of distant supernova
redshifts~\cite{Riess:1998cb,Perlmutter:1998np}.  Detailed studies of
anisotropies in the cosmic microwave background
radiation~\cite{Bennett:2003bz} and broad surveys of large-scale
structure~\cite{Cole:2005sx,Tegmark:2003ud} have deepened and broadened
the initial evidence. Remarkably, most of the stuff of the Universe 
appears to lie outside our quotidian experience, and has not yet been 
detected in the laboratory. 

The spottiness of the fossil record that we read in distance-redshift
correlations, microwave anisotropies, and large-scale structure leaves
much room for interpretation~\cite{Bridle:2003yz}.  The most economical
description of the cosmological parameters attributes the
negative-pressure (``dark energy'') component to a cosmological
constant in Einstein's equation---an omnipresent and invariable vacuum
energy density that assumes a greater importance as the Universe
expands~\cite{Krauss:1995yb}.  On this picture, we are entering a final
inflationary epoch in which the Universe will grow so quickly as to be
essentially empty of matter.  A dynamical alternative attributes the
accelerated expansion to a cosmic scalar field that changes with time
and varies across space, slowly approaching its ground
state~\cite{Caldwell:1997ii,Wang:1999fa}.  Such quintessence models, as they are
called, admit a broad variety of future Universes.

Neither the cosmological constant interpretation nor the cosmic-scalar
picture has a ready explanation for the rough balance between matter
and vacuum energy at this moment in cosmic history.  The cosmic
coincidence problem---the ``why now?'' question--has stimulated
speculations that range from anthropic
rationalizations~\cite{Susskind:2003kw} to cyclic
cosmologies~\cite{Steinhardt:2002ih}.

Recently, we investigated the possibility that the physical
characteristics of the vacuum energy might vary with time, specifically
with the number of $e$-foldings of the scale factor $a$ of the
Universe~\cite{Barenboim:2004kz}.  We showed that the simple
\textit{Ansatz,} 
\begin{equation}
    w_{v}(a) = -\cos(\ln a)\;,
    \label{eq:oureos}
\end{equation}
for the vacuum-energy equation of state is compatible with existing
observations and offers an intriguing response to the ``why now?''
problem.  (We choose a phase implicit in the form \eref{eq:oureos}  to
match the inference that $w_{v0} \approx -1$ in the current Universe.)
The cosmic expansion of the ensuing \textit{undulant universe} is
characterized by alternating eras of acceleration and deceleration. 
Because over one period the equation of state \eref{eq:oureos} 
averages to zero (the equation of state of
pressureless matter), 
the vacuum energy density tracks the matter density on average. In 
this way, 
the cosmic coincidence problem is resolved.  
The future evolution of the undulant universe is similar in broad
outline to that of a matter-dominated universe at critical density, for
which $a \propto t^{2/3}$, where $t$ measures the age of the Universe.

The oscillatory equation of state \eref{eq:oureos} could represent, for
instance, the effective description of a dynamical cosmic field (fluid)
present since very early times.  In that spirit, Barenboim and
Lykken~\cite{Barenboim:2005np} have extended the undulant universe
notion, taking a new approach to quintessential inflation that not only
yields inflation but also offers a dark energy candidate.  Both
features emerge from the evolution of a single scalar field in a
potential with oscillatory and exponential behavior.  (The potential of
Ref.~\cite{Barenboim:2005np} resembles the motion of a
Slinky\textregistered\ spring toy descending a staircase, so the
paradigm is called slinky inflation.)  The vacuum-energy equation of
state that emerges from their potential has the form
\begin{equation}
   w_{v}(a) = -\cos(b \ln a)\;,
   \label{eq:oureos2}
\end{equation} 
a simple generalization of \Eref{eq:oureos}, in which the dimensionless
parameter $b$ controls the frequency of inflationary epochs. They 
present an illustrative example with $b = \cfrac{1}{7}$ that reproduces 
measured energy densities in the present universe and produces 
the requisite inflation. After the end of the radiation-dominated era, 
indeed, back to $a \approx 10^{-8}$, the characteristics of slinky 
inflation are essentially identical to those of an undulant cosmology 
specified by the equation of state \eref{eq:oureos2}. At earlier 
times, it is necessary to solve the coupled differential equations 
for the radiation, matter, and vacuum-energy densities given by 
Equation~(12) 
of Ref.~\cite{Barenboim:2005np}.

It is interesting to observe that equations
of state involving the functional form $\cos(\ln{a})$ but passing
through $w = -1$ have been explored, to a different end, in a number of
recent 
papers~\cite{Feng:2004ff,Guo:2004fq,Xia:2004rw,Wei:2005nw,Lazkoz:2005sp}.  

In this paper we catalogue the observational tests to which we have
subjected the undulant universe defined by the simple equation of state
\eref{eq:oureos}.  We further explore the bounds that can be placed on
the frequency parameter $b$ by requiring consistency with current
observations.  In the limit of small values of $b \to 0$, the equation
of state \eref{eq:oureos2} approaches the cosmological-constant value,
$w \approx -1$.  In that limit, the periodic equation of state merely
reproduces the success of the cosmological constant + Cold Dark Matter
($\Lambda$CDM) hypothesis (and offers no insight into the ``why now''
problem).  More to the point, we find that a periodic equation of state
with $0.6 \ltap b \ltap 2$, which responds to the cosmic coincidence
problem, is in comfortable agreement with all observations, save for
the demands of structure formation as expressed through linear
evolution of density perturbations.  We shall explore this
vulnerability.

We examine ways in which future observations might further constrain,
or rule out, variants of the undulant universe.  The fact that an
alternative so different from the \lcdm\ hypothesis is compatible with
observations makes it important to pursue the discovery of the
accelerating expansion on multiple fronts.  In addition to seeking to
characterize the dark energy through its equation of state in the
recent past, it is highly desirable to probe the state of the Universe 
at epochs for which we have not yet learned to read the fossil record. 
It is premature to converge on a single hypothesis.

\section{The essence of the undulant universe}
Let us recapitulate the main elements of an evolving universe.  The
expansion of the universe is determined by the Friedmann equation,
\begin{equation}
    H^{2} \equiv ({\dot{R}}/{R})^{2} = {8\pi 
    G_{\mathrm{N}}\rho}/{3} - {k}/{R^{2}} + {\Lambda}/{3}\;,
    \label{eq:friedmann}
\end{equation}
where $H$ is the Hubble parameter, $R$ is the cosmological  
scale factor, $G_{\mathrm{N}}$ is Newton's constant, $\rho$ is the 
energy density, $k = 
(+1,0,-1)$ is the curvature constant, and $\Lambda$ is a possible 
cosmological constant. If $\Lambda = 0$, the 
curvature constant determines cosmic destiny. For $k=+1$ (closed Universe), 
the Universe recollapses in finite time; for $k=0$ (flat) and $k=-1$ 
(open), the Universe expands without limit. It is convenient to define the dimensionless 
scale factor, $a = R/R_{0}$, where the subscript $0$ denotes the 
value at the current epoch. The critical density, defined from 
\eref{eq:friedmann}, is $\rho_{c} = {3H^{2}}/{8\pi G_{\mathrm{N}}}$.
The dimensionless cosmological density parameter is defined relative to the 
critical density as $\Omega_{\mathrm{tot}} = \rho/\rho_{c}$ at any 
epoch. 
We express the rate of change of the Hubble parameter through the 
deceleration parameter,
\begin{equation}
    q \equiv - \frac{1}{H^{2}} \, \frac{\ddot{R}}{R} = 
    \frac{\Lambda}{3H^{2}} - \frac{4\pi 
    G_{\mathrm{N}}}{3H^{2}}\,(\rho + 3p)\;,
    \label{eq:decel}
\end{equation}
where $p$ is the isotropic pressure. If we define 
$\Lambda = 4\pi G_{\mathrm{N}}\rho_{\Lambda}$ and introduce the 
equation of state $w_{i} = p_{i}/\rho_{i}$ for any component of the 
universe, we can recast the deceleration parameter as
\begin{equation}
    q = \cfrac{1}{2}\sum_{i}\Omega_{i}(1 + 3 w_{i}) = \cfrac{1}{2} 
    \left(\Omega_{\mathrm{tot}} + 3 \sum_{i}\Omega_{i}w_{i} \right)\;.
    \label{eq:decel2}
\end{equation}
The equation of state of pressureless matter is $w_{m}=0$, and that of 
radiation is $w_{r}= \cfrac{1}{3}$. We see by inspection of 
\eref{eq:decel} that $w_{\Lambda} = -1$. Note that the deceleration 
parameter is defined to be positive if the rate of expansion is slowing; 
for the reference case (SCDM) of a matter-dominated universe at critical 
density ($\Omega_{\mathrm{tot}} = 1, \Lambda = 0$), $q = 
\cfrac{1}{2}$.  

The $\Lambda$CDM proposal is attractive for its simplicity, and it
agrees well with all observations.  It does appear to bear a burden of
unnaturalness, however.  While $\Lambda$CDM predicts that
$\Omega_{\Lambda} \approx \Omega_{m}$ at some point during the
evolution of the Universe, it offers no explanation for the curious
circumstance that the balance occurs at the current epoch---and no
other---in the history of the Universe.

Let us analyze what happens if the physical characteristics of the
vacuum energy vary periodically with the number of $e$-foldings of the
scale factor according to the equation of state \eref{eq:oureos} or the
generalization \eref{eq:oureos2}.  In the numerical examples that
follow, we assign the vacuum energy a weight $\Omega_{v0}=0.7$, in line
with observations, and take $\Omega_{m0}=0.3$ and $\Omega_{r0} =
4.63\times10^{-5}$.  The
present-day expansion rate is $H_{0} = 100\,h\km\s^{-1}\mpc^{-1}$, with
$h = 0.71^{+0.04}_{-0.03}$~\cite{Eidelman:2004wy}.

\begin{figure}
   \centerline{\includegraphics[width=10cm]{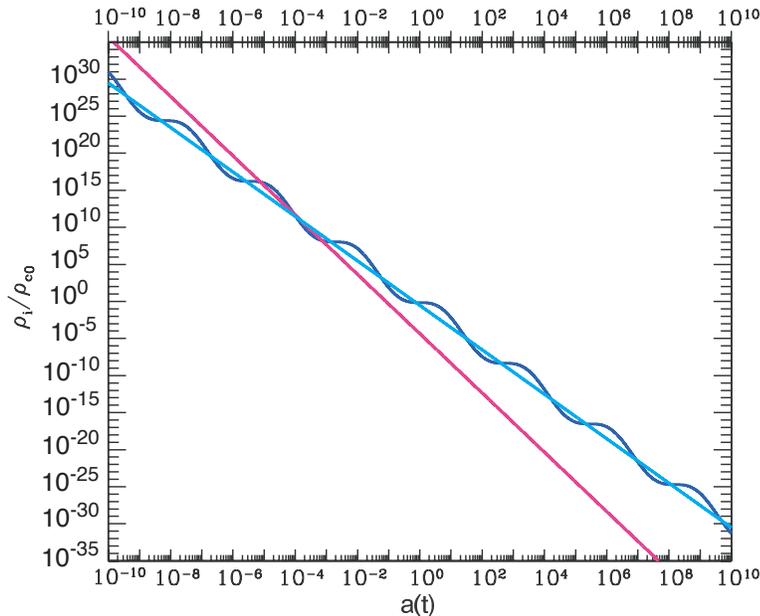}}
   \vspace*{-12pt}
\caption{Evolution of the matter (straight cyan line), radiation
(magenta, steepest line), and vacuum (undulating blue line) energy
densities in the undulant universe, normalized to the critical density
$\rho_{i}/\rho_{c0}$, versus the scale factor $a(t)$.}
\label{fig:fish}
\end{figure}

In the general case \eref{eq:oureos2}, these are given in
terms of the present normalized densities as
$\rho_{m}/\rho_{c0} = \Omega_{m0}/a^{3}$, $\rho_{r}/\rho_{c0} = 
\Omega_{r0}/a^{4}$, and $\rho_{v}/\rho_{c0} = g(a)\Omega_{v0}/a^{3}$, 
where 
\begin{equation}
    g(a) =  e^{3\int_a^1 d{a^\prime}\,w(a^\prime)/a^{\prime} }
  =   \exp{[\cfrac{3}{b} \sin(b\ln{a})]}\;.
    \label{eq:gofa}
\end{equation}
We plot in \Fref{fig:fish}
the normalized energy densities of matter, radiation, and vacuum energy
as functions of the scale parameter $a$, for the undulant universe case
of $b = 1$.  Looking back in time to the epoch of big-bang
nucleosynthesis at $a \approx 10^{-10}$, and forward to $a = 10^{+10}$,
we see that the vacuum energy density crosses the matter density every
$\pi$ $e$-foldings of the scale factor.  If we require the vacuum
component to have negative pressure, a situation similar to the present
occurs every $2\pi$ $e$-foldings.  Periodically dominant dark energy
is in the spirit of 
Refs.~\cite{Dodelson:2001fq,Griest:2002cu,Sahni:1999qe,Ferrer:2005hr}. These
regular crossings stand in sharp contrast to the $\Lambda$CDM
cosmology, in which $\Omega_{v} \approx \Omega_{m}$ only in the
current epoch.

\begin{figure}[tb]
   \centerline{\includegraphics[width=6.5cm]{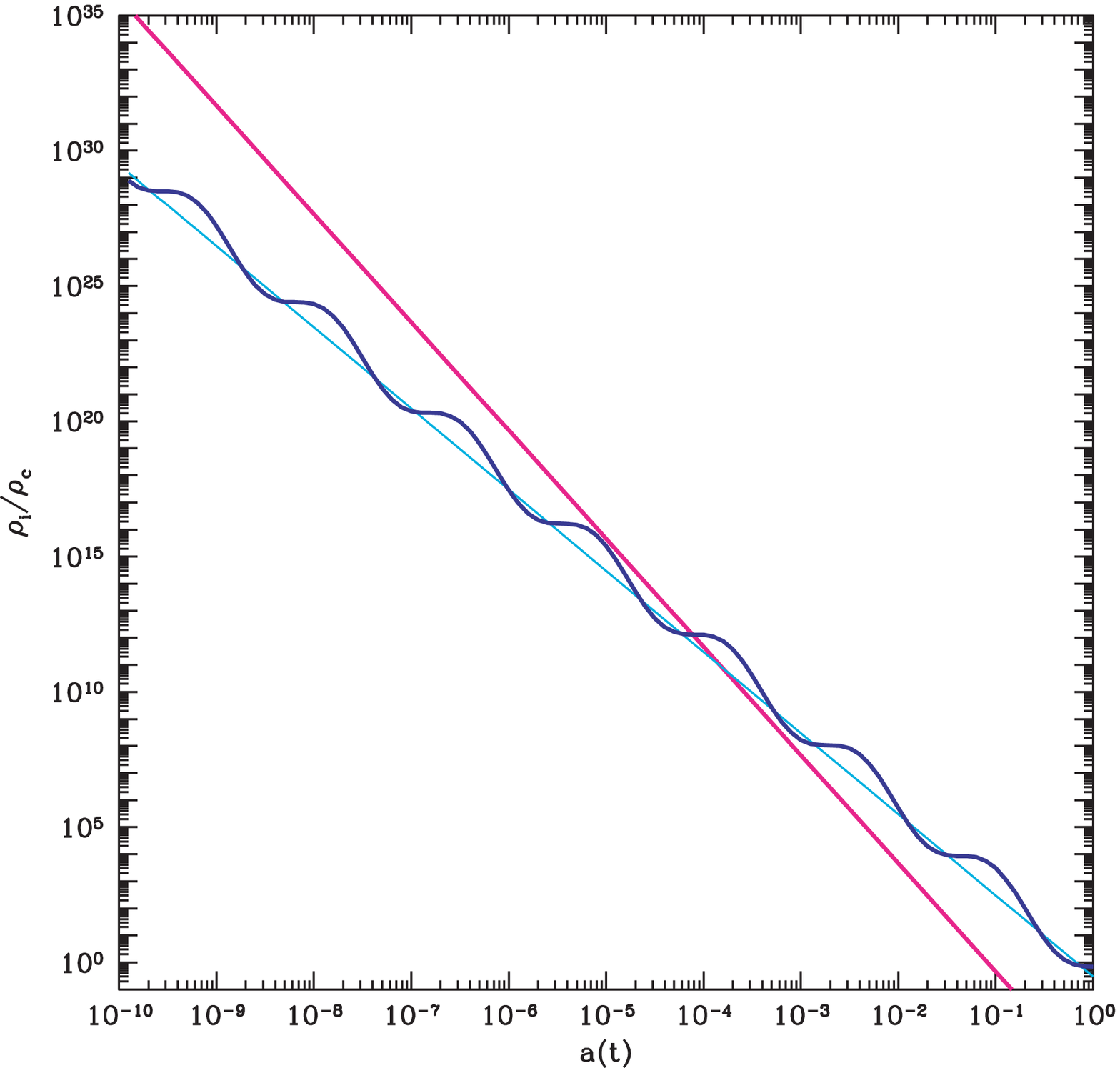}\quad
   \includegraphics[width=6.5cm]{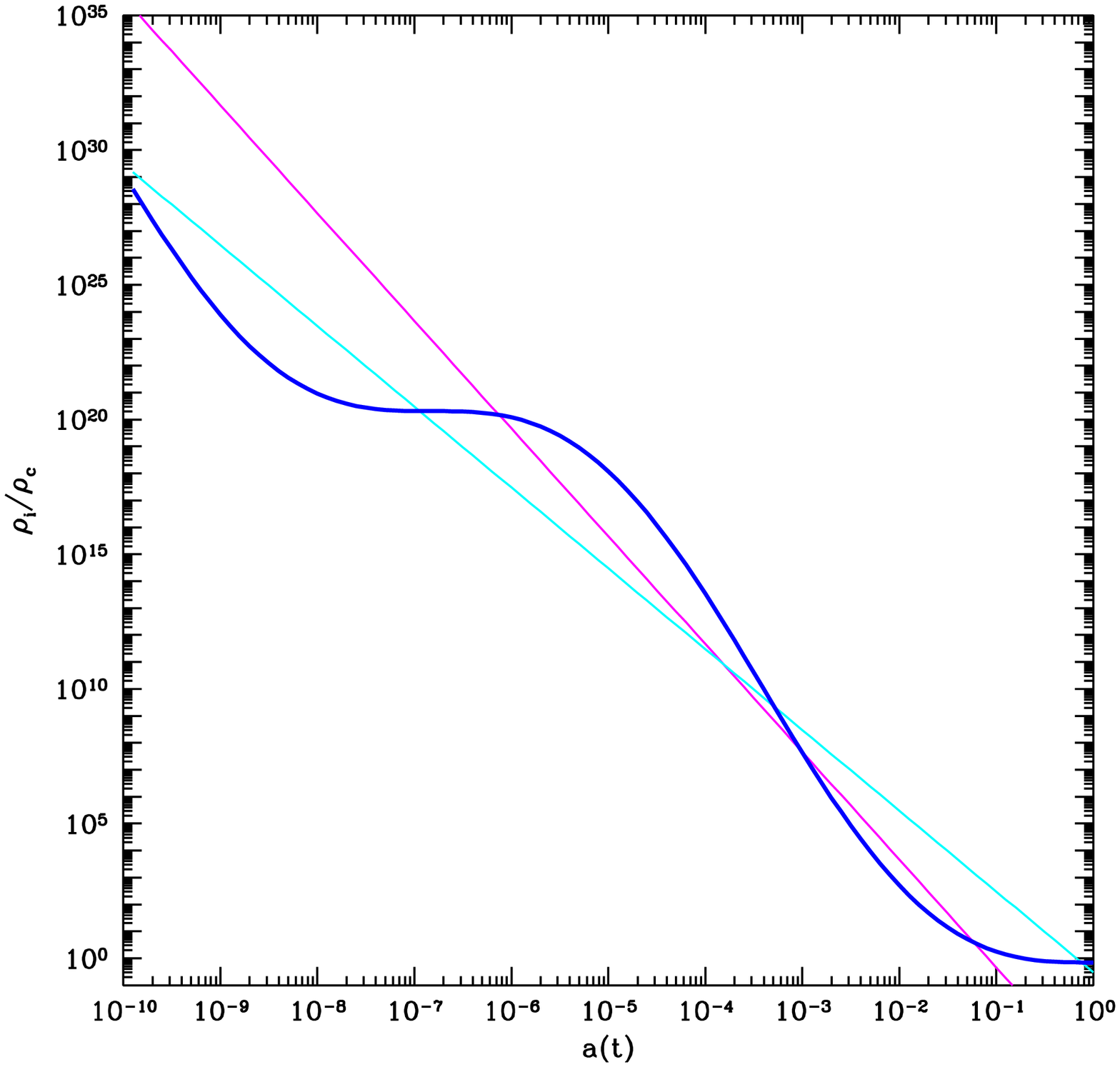}}
   \vspace*{6pt}
   \centerline{\includegraphics[width=6.5cm]{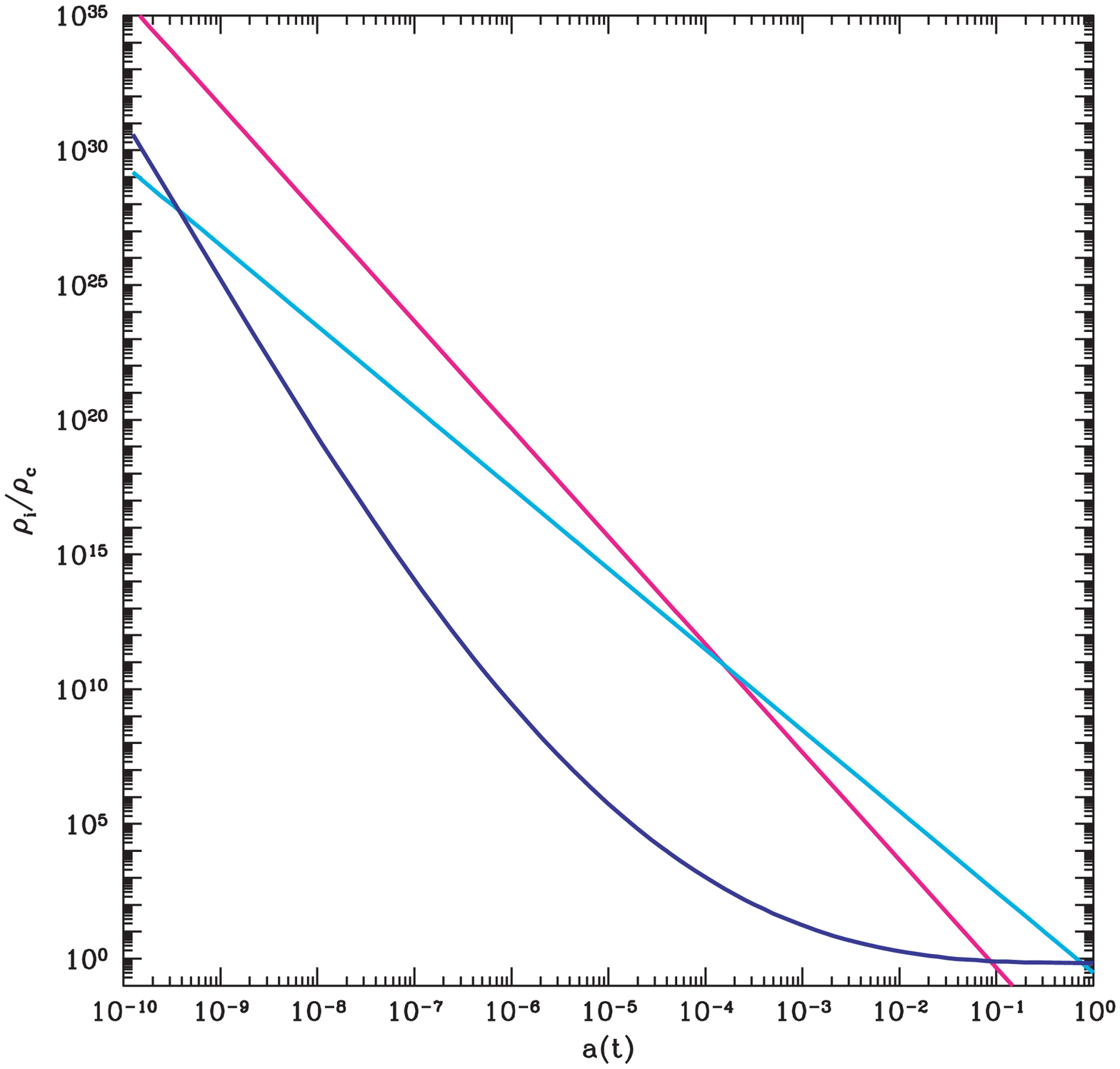}\quad
   \includegraphics[width=6.5cm]{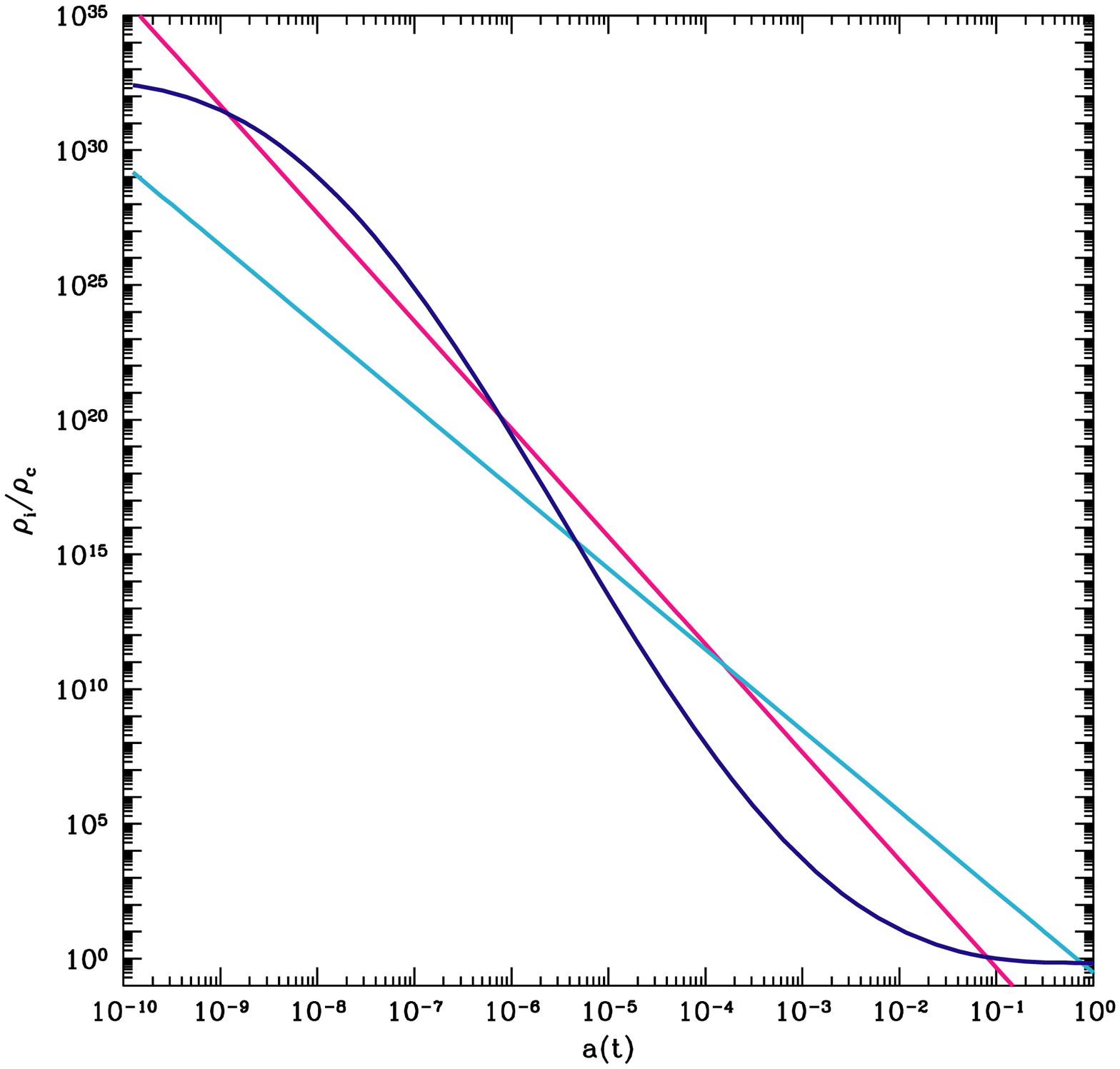}}
\vspace*{-6pt}
\caption{Evolution from $a = 10^{-10}$ to the present 
of the matter (straight cyan line), radiation
(magenta, steepest line), and vacuum (undulating blue line) energy
densities in the variants of the undulant universe, normalized to the critical density
$\rho_{i}/\rho_{c0}$, for different frequency parameters $b$. 
Clockwise from top left, the plots correspond to $b = 2$, $b = 
\cfrac{2}{5}$, $b = \cfrac{1}{4}$, and $b = \cfrac{1}{7}$.}
\label{fig:fishies}
\end{figure}

The presence of the exponential factor $e^{3/b}$ in \eref{eq:gofa}
raises the possibility that for small values of $b$, the excursions in
the vacuum energy density about the matter density may be amplified to
unacceptable---or at least highly nonstandard---levels.  We plot in
\Fref{fig:fishies} the histories of the Universe that result from
the choices $b = 2, \cfrac{2}{5}, \cfrac{1}{4}, \cfrac{1}{7}$.  For $b
= 2$ (top left panel), the oscillations are more frequent, and smaller,
than those for the canonical choice $b = 1$.  For $b = \cfrac{2}{5}$
(top right), the oscillations are larger, but less frequent, than for
the default undulant Universe.  The radiation-dominated era ends a
little earlier than in the standard cosmology, but nothing is overtly
wrong.  The case $b = \cfrac{1}{4}$ (bottom right) is highly
problematic, however: the first positive excursion occurs in the era of
big-bang nucleosynthesis (BBN), and would imply a large vacuum-energy
component at that time.  [The equation of state is $w_{v} = -0.86,
-0.45, +0.11$ for $a = 10^{-10}, 10^{-9}, 10^{-8}$.] At still smaller 
values of $b$, including the case $b =\cfrac{1}{7}$ (bottom left), the 
first positive excursion occurs well before BBN, and would not raise 
any obvious problems.

\begin{figure}
\centerline{\includegraphics[width=10cm]{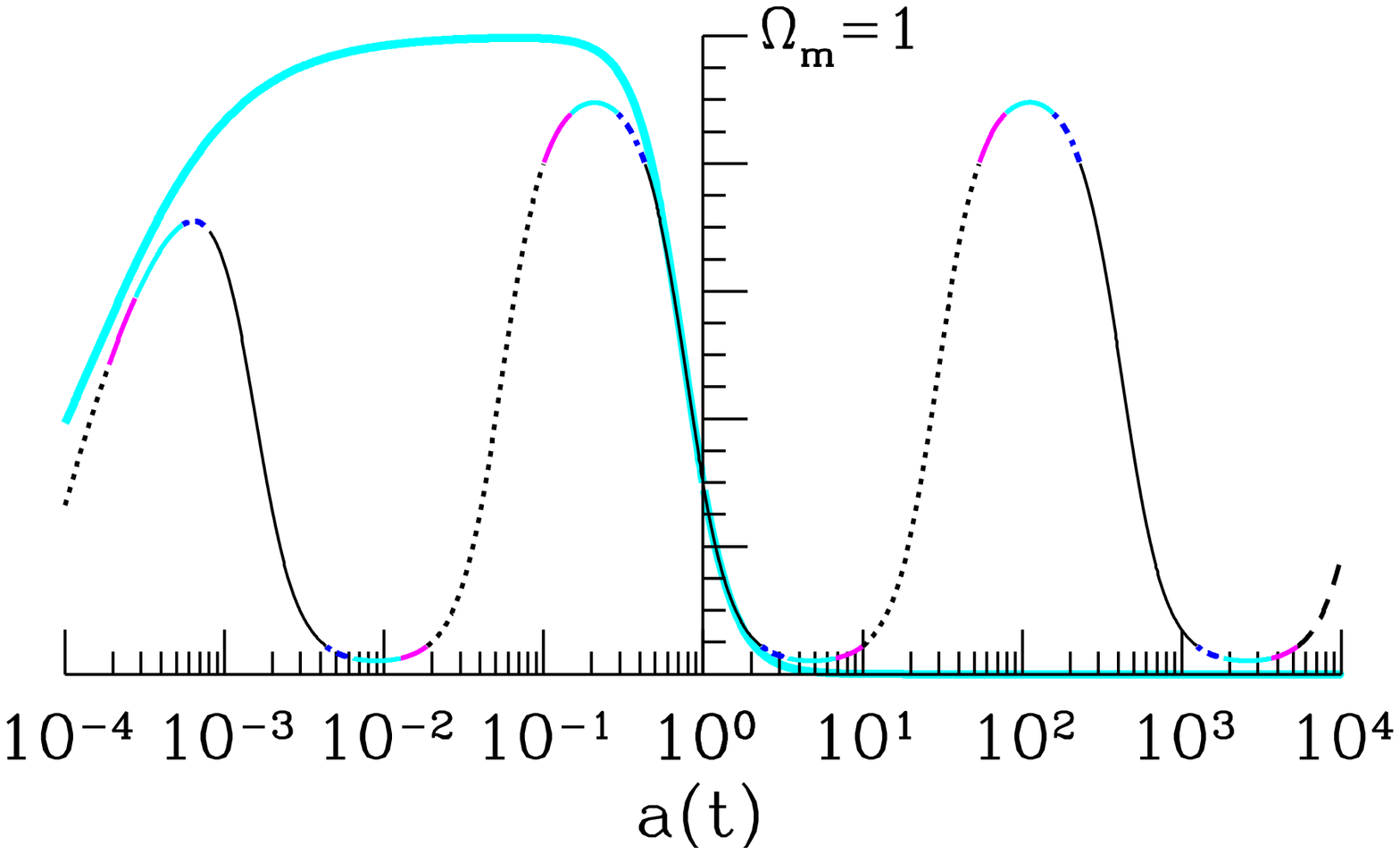}}
\vspace*{6pt}
\centerline{\includegraphics[width=10cm]{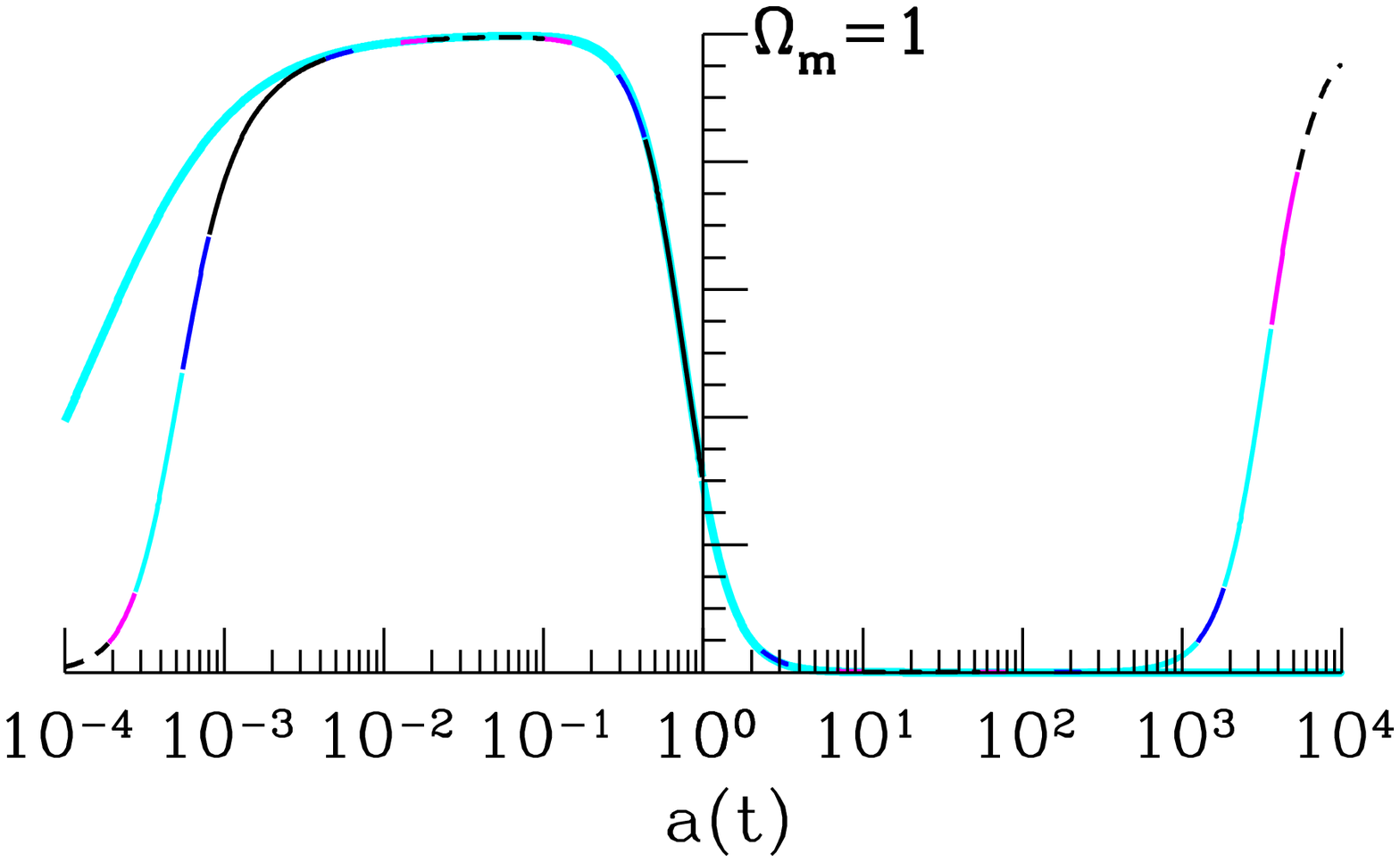}}
\caption{{Evolution of the matter fraction as a function of scale
factor for the $\Lambda$CDM model (thick cyan curve) and for the
undulant universe described by \eref{eq:oureos} (oscillatory curve).
At different eras, the vacuum fluid emulates radiation (magenta),
matter (cyan), small negative values of $w$ (blue, dash-dotted), dark
energy (black solid), and a scalar field in the kination regime
(dotted).} Top panel: undulant universe of \eref{eq:oureos}; bottom 
panel: generalized undulant universe of \eref{eq:oureos2}, with 
frequency parameter $b = \cfrac{2}{5}$. }
\label{fig:fig3}
\end{figure}

Although the matter density in the Universe scales smoothly as $a^{-3}$
with respect to today's critical density, the matter fraction of
mass-energy at any moment, $\rho_{m}(a)/\rho_{\mathrm{tot}(a)}$, is
influenced by the evolution of the rest of the portfolio.  In the case
of the $\Lambda$CDM picture, the matter density was dominant in the
recent past and will be totally negligible in the near future, because
the vacuum energy density due to the cosmological constant is
independent of scale factor.  The resulting evolution is shown by the
thick curve in \Fref{fig:fig3}: matter takes over from radiation as the
dominant component as the scale factor increases through $a = 10^{-4}$.
At recent times, corresponding to $a \gtap 10^{-1}$, matter is
supplanted by vacuum energy once and for all, and quickly becomes a
negligible fraction of the mass-energy budget.

In the undulant cosmology implied by the equation of state
\eref{eq:oureos}, the matter fraction is periodically dominant, not
only at the end of the radiation era, but also at regular intervals in
the future.  The approximate cycles repeat every $2\pi/b$ $e$-foldings,
where $b$ is the frequency parameter in \eref{eq:oureos2}.  The
changing equation of state of the vacuum energy is 
encoded in the color
and texture of the undulating curve in \Fref{fig:fig3}.
The dotted black  line corresponds to $1> w> \cfrac{2}{3}$;
magenta to  $\cfrac{2}{3} >w>\cfrac{1}{3}$;
cyan to $\cfrac{1}{3} >w>-\cfrac{1}{3}$;  
dash-dotted blue to $-\cfrac{1}{3} > w >-\cfrac{2}{3}$;
solid black to $-\cfrac{2}{3} > w > -1$.

The Hubble parameter  corresponding to scale factor $a$ is  given by
\begin{equation}
    H(a) = H_0 \sqrt{ \frac{\Omega_m}{a^{3}} + \frac{g(a)
    \Omega_v}{a^{3}} +\frac{\Omega_r}{a^{4}} }\;.
    \label{eqn:hubble}
\end{equation}
In the undulant universe (with $b = 1$), the current age of the
universe, $t_{0} = \int_0^1 da/H(a) a$, is $13.04\gyr$, to be compared
with $13.46\gyr$ in the $\Lambda$CDM model.  By calculating the time to
reach a given scale factor, we can determine the history and future of
the universe.  During the radiation-dominated era, which corresponds to
$a \ltap 10^{-5}$, $a(t) \propto t^{1/2}$; when matter dominates, $a(t)
\propto t^{2/3}$.

\begin{figure}
\centerline{\includegraphics[width=7cm]{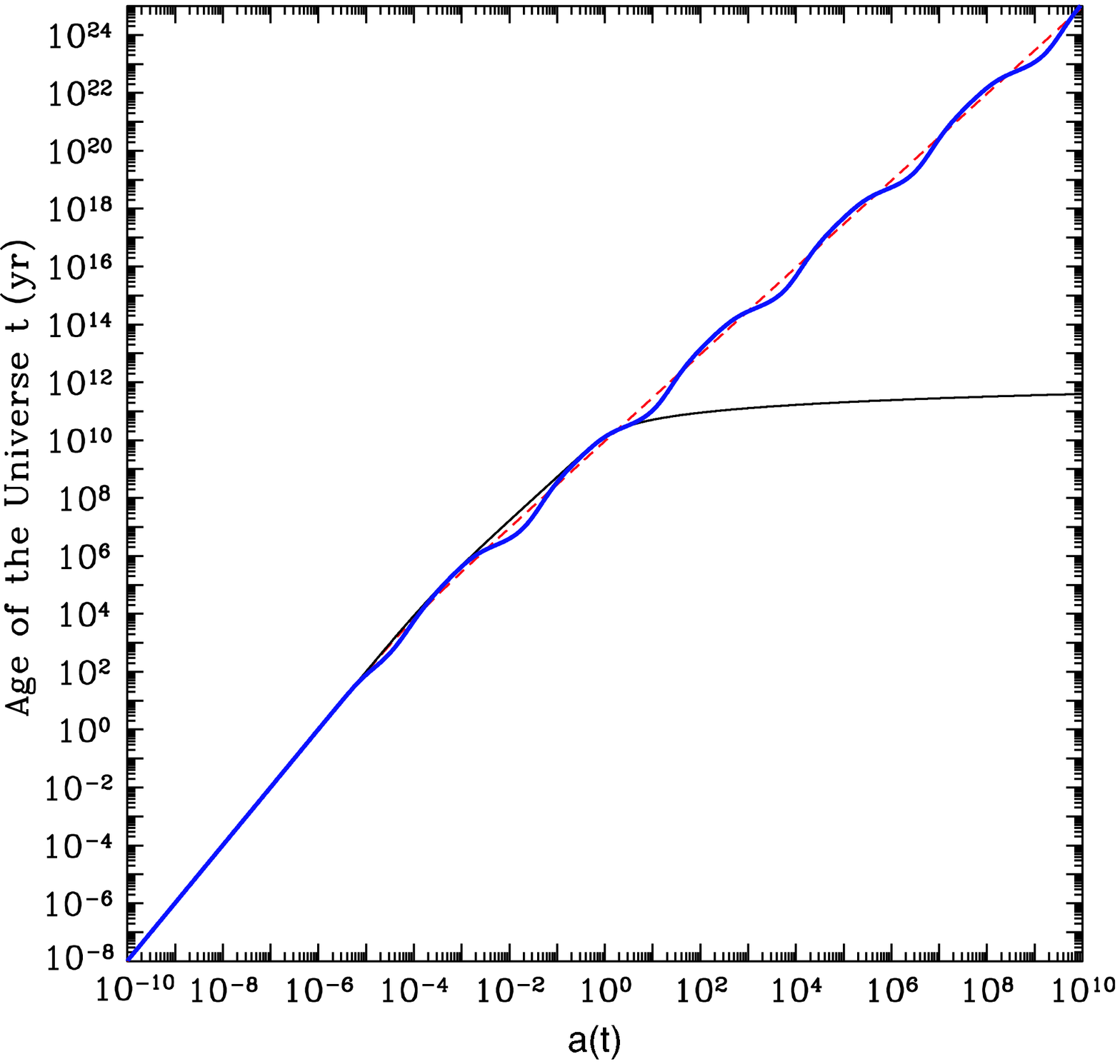}
\quad\includegraphics[width=7cm]{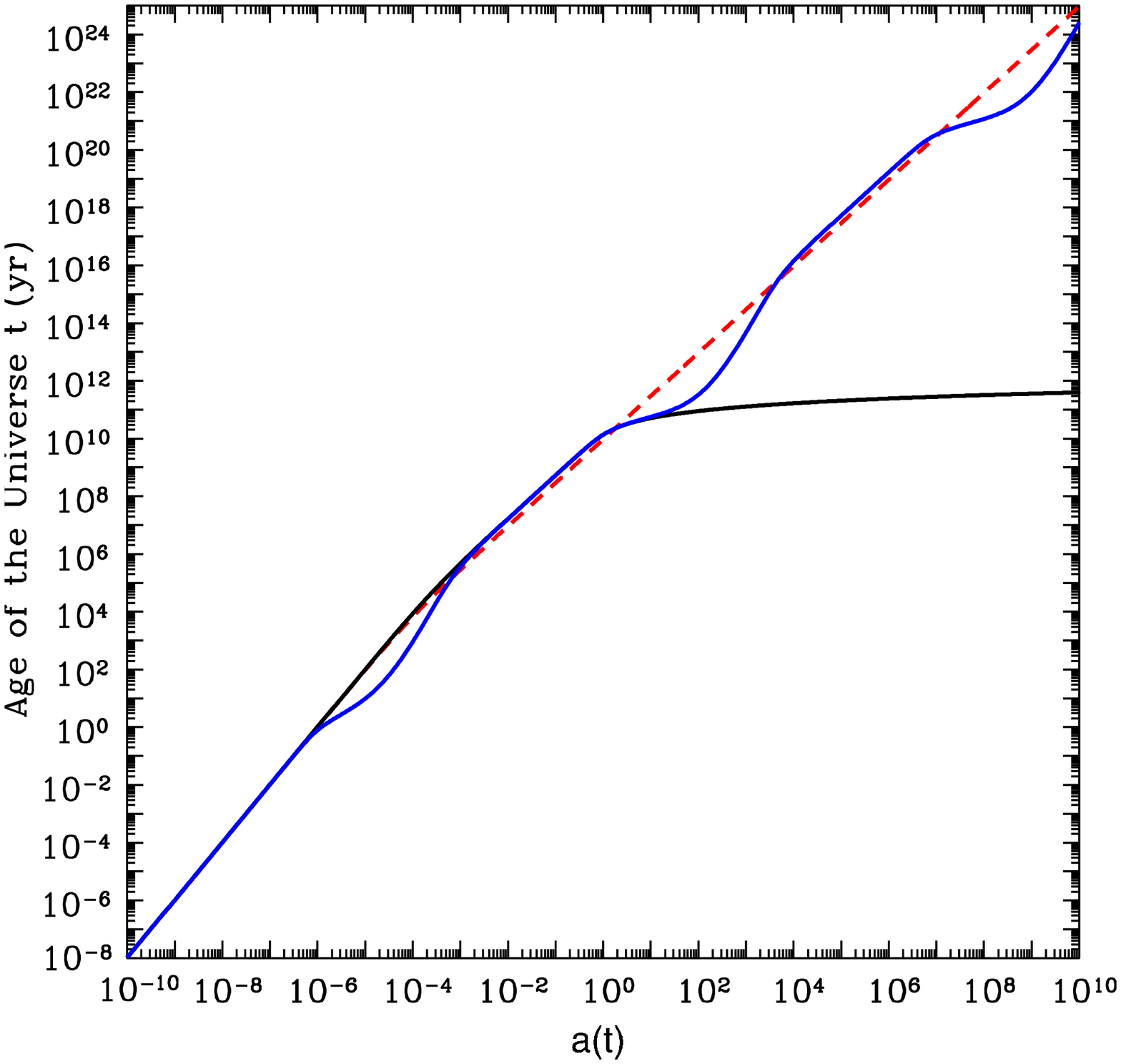}}
\caption{Evolution of the scale factor $a(t)$ in three cosmologies: the
canonical $\Lambda$CDM model (thin black line); a critical universe
(SCDM model) with $\Omega_{m} = 1$ (dashed red line); and (left panel)
the periodic equation of state \eref{eq:oureos} or (right panel) the 
periodic equation of state \eref{eq:oureos2} with $b=0.4$ (thick blue line).  }
\label{fig:scalefactor}
\end{figure}

We show the time dependence of the scale factor $a(t)$ for three
cosmologies in \Fref{fig:scalefactor}. 
The dashed (red) line corresponds to the ``standard cold dark 
matter'' (SCDM) cosmology that was canonical before the discovery of 
the accelerating universe. The thin solid (black) line shows the 
$\Lambda$CDM cosmology, in which the present epoch marks the beginning of 
a final inflationary period that leads to an empty universe in which 
matter is a negligible component. The heavy (blue) line shows the 
prediction that follows from \Eref{eq:oureos}. In the recent past, the 
periodic equation of state matches the behavior of the $\Lambda$CDM 
cosmology, but in the future it undulates about the SCDM prediction. 

\begin{figure}
\centerline{\includegraphics[width=8.5cm]{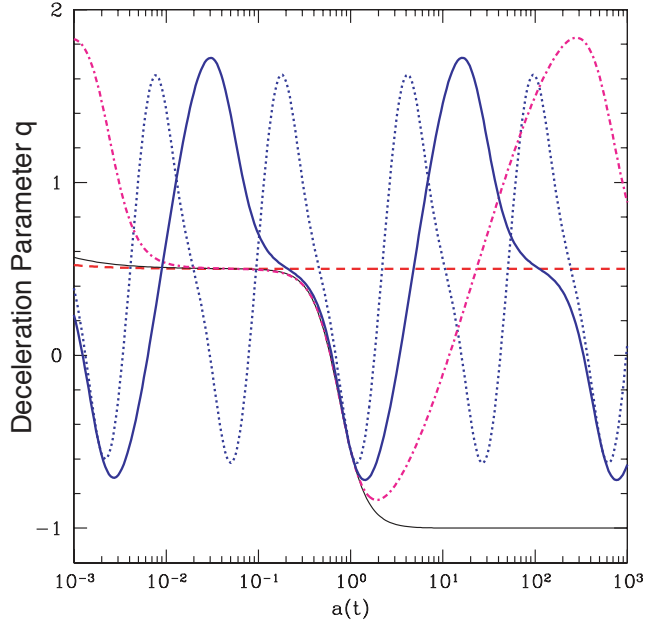}}
\caption{The deceleration parameter as defined in \Eref{eq:decel2}
for the undulant universe of \eref{eq:oureos} (thick blue line),
$\Lambda$CDM model (thin black line) and SCDM model (dashed red line).
Variants of the undulant universe given by the generalized equation of
state \eref{eq:oureos2} are shown as the dot-dashed line, for $b =
\cfrac{1}{2}$, and the dotted line, for $b = 2$.}
\label{fig:decel}
\end{figure}

The alternating periods of acceleration and deceleration that
characterize the expansion of \textit{undulant universes} are
signaled by the deceleration parameter in \Fref{fig:decel}.  For scale
factors
$a$ between $0.1$ and $1$, the periodic equation of state 
\eref{eq:oureos} tracks the
behavior of the $\Lambda$CDM cosmology.  For the generalized form 
\eref{eq:oureos2}, undulant cosmologies trace the $\Lambda$CDM curve 
over greater or smaller ranges of the scale factor for values of the 
frequency parameter $b$ that are smaller or greater than unity. 
In contrast to the $\Lambda$CDM universe, which
about to enter a final inflationary era of sustained acceleration, the
average behavior of the undulant universes tracks that of SCDM.

Let us take a closer look at the ``why now?''  problem.  In the
$\Lambda$CDM picture, there is literally no (other) time like the
present, but in undulant cosmologies the matter--dark-energy
coincidence is a more or less typical circumstance.  We will need to
define a statistical measure to assess how typical is the present state
of the Universe.  It is informative to distribute a large number
$\mathcal{N}$ of points randomly in $\ln(a)$, excluding a slice that
contains the present Universe, and to ask what fraction correspond to a
balance between matter and dark energy similar to what we observe
today.  We define the current epoch by the condition $0.7 < a < 1.6$, 
and say that the current situation is matched, provided that 
\begin{eqnarray}
    0.1 \le & \frac{\Omega_{m}}{\Omega_{m}+\Omega_{v}} & \le 0.5\;, 
    \hbox{ and}
    \nonumber  \\
    0.5 \le & \frac{\Omega_{v}}{\Omega_{m}+\Omega_{v}} & \le 0.9\;, 
    \hbox{ with } w_{v} < -0.7\;.
    \label{eq:success}
\end{eqnarray}
[For scale factors $a \gtap 10^{-5}$, radiation contributes negligibly 
to the energy portfolio of the Universe, so that 
$\Omega_{m}+\Omega_{v} \approx 1$.]

We examine two intervals in the scale factor.  First, we consider
responses to the cosmic coincidence question since the era of radiation
dominance $a > 10^{-5}$, when the Universe was about 94 years old, and
a corresponding step into the future, extending to $a = 10^{5}$.  We
tally the number of successful throws of $\ln(a)$, out of $\mathcal{N}
= 10^{6}$ trials, and report the probability of success in \Tref{tbl:stats}.
\begin{table}[tbp]
    \centering
    \caption{Probability, in percent, that conditions approximate the current 
    matter--dark-energy energy balance of 
    the Universe, according to the criteria explained in the text, 
    for various values of the frequency parameter $b$ of 
    \eref{eq:oureos2}.}
    \vspace*{12pt}
\begin{center}
	\begin{tabular}{|c||c|c||c|c|}
	    \hline
	    $b$ & $10^{-5}< a < 0.7$  & $1.6 < a < 10^{5}$
	& $10^{-20} < a < 0.7$ & $1.6 < a < 10^{20}$  \\
	    \hline
	    $0$ ($\Lambda$CDM) & 0 & 0 & 0 & 0  \\
	    \hline
	    $0.1$ & $0$ & 0 & 0 & 0  \\
	    \hline
	    $0.125$ &  $0$ & 0 & 0 & 0 \\
	    \hline
	    $1/7$ & $0$ & 0 & $1.5$ & $1.6$  \\
	    \hline
	    $0.25$ & $0$ & 0 & $1.6$ & $1.6$  \\
	    \hline
	    $0.4$ & $0$ & 0 & $3.2$ & $3.2$  \\
	    \hline
	    $0.5$ & $0$ & 0 & $4.8$ & $4.8$  \\
	    \hline
	    $0.6$ & $6.4$ & $6.4$ & $6.4$ & $6.4$  \\
	    \hline
	    $0.7$ & $6.4$ & $6.4$ & $7.8$ & $8.0$  \\
	    \hline
	    $0.9$ & $6.5$ & $6.5$ & $9.8$ & $9.8$  \\
	    \hline
	    $1$ & $6.6$ & $6.7$ & $11.3$ & $11.5$  \\
	    \hline
	    $2$ & $18.2$ & $18.2$ & $21.1$ & $21.2$ \\
	    \hline
	    $5$ & $24.9$ & $24.9$ & $24.8$ & $25.3$  \\
	    \hline
\end{tabular}
\end{center}
    \label{tbl:stats}
\end{table}
In the undulant universe defined by $b = 1$~\cite{Barenboim:2004kz}, 
success comes about one time in sixteen, both in the past and in the 
future. The current conditions are in that sense typical of the undulant 
universe. This statistical conclusion squares with the behavior we 
observed in \Fref{fig:fish}. As the frequency of undulations 
increases, conditions like those in the current Universe become 
commonplace. On the other hand, for values of the frequency parameter 
$b \ltap 0.6$, the present coincidence of mass and dark energy is the 
only one to occur in the range $10^{-5} \le a \le 10^{5}$. For the 
range $10^{-7} \le a \le 10^{7}$, values of $b \ltap \cfrac{3}{8}$ 
yield only the present coincidence.

Expanding the range over we look for cosmic coincidences to the
interval $10^{-20} \le a \le 10^{20}$, we find that the undulant 
universe with $b = \cfrac{1}{7}$ implies one coincidence in the
past and one in the future, but that models in which $b \ltap
\cfrac{1}{7}$ imply no coincidences other than the present one, which
is included by design. It is fair to question whether a 
matter--dark-energy coincidence in the radiation-dominated universe 
($a \ltap 10^{-5}$) is of any moment. The tracking behavior of slinky 
inflation, in contrast, ensures that radiation is not uniformly 
dominant from the big bang to $a \approx 10^{-5}$, and opens the way 
to early coincidences that are more generally meaningful.

\section{Tests of Undulant Cosmologies}
The principal constraints on cosmological models arise from 
observational knowledge of the conditions that prevailed at the time 
of big-bang nucleosynthesis, studies of the 
cosmic microwave background that look back to the surface of last 
scattering ($a \approx 10^{-3}$), and measurements of the deceleration 
parameter in supernova redshift surveys near the present epoch, 
extending to redshifts $z \approx 2$. The growth of large-scale 
structure also exhibits some sensitivity to the cosmic equation of 
state, through a tension between the attraction of gravitational 
stability and the dynamical friction of the expansion, but the 
standard treatment involves only linear 
perturbations~\cite{Linder:2003dr}. We shall discuss these in turn.

\subsection{Big-bang nucleosynthesis \label{subsec:bbn}}
The presence of a dark energy field at early times alters the expansion
rate of the Universe, changing the ratio of neutrons to protons at
freeze-out and modifying the predicted abundances for light elements.
At those early times ($a \ltap 10^{-9}$) the vacuum-energy density
implied by the undulant universe \eref{eq:oureos} is utterly
negligible in comparison to radiation, as we saw in
\Fref{fig:fish}, and so its influence on big-bang nucleosynthesis
will be imperceptible. We find, for example, $\Omega_{v}(a = 10^{-10}) 
\approx 2 \times 10^{-5}$, and $\Omega_{v}(a = 10^{-9})  \approx  
10^{-6}$.  

Bean, Hansen, and Melchiorri~\cite{Bean:2001wt} have reported that the
$^{4}\mathrm{He}$ mass fraction and the deuterium-to-hydrogen ratio do
not favor the presence of a dark-energy component.  In a class of
quintessence models, they determine the bound $\Omega_{v} < 0.045$ at
$2\sigma$.  We can take the resulting bound as a reasonable upper limit
for the more general case of the periodic equation of state
\eref{eq:oureos2}.  This bound is not threatened for $0.5 \ltap b
\ltap 2$.  However, the exponential factor $e^{3/b}$ in the expression
\eref{eq:gofa} for $g(a)$ can grow so large, for small values of the
frequency parameter $b$, that the energy budget of the Universe is
altered dramatically just at the BBN era.  We see this effect in the
$b=1/4$ (bottom right) panel of \Fref{fig:fishies}.  The
large value $\Omega_{v}^{b=1/4}(a = 10^{-9}) \approx 0.4$ is clear
cause for concern.  Smaller values, including $b = \cfrac{1}{7}$, for
which a large positive excursion of the vacuum-energy density occurs
well before BBN, do not provoke any concern on this score.
 
\subsection{Power spectrum}
\begin{figure}
\centerline{\includegraphics[width=8.5cm]{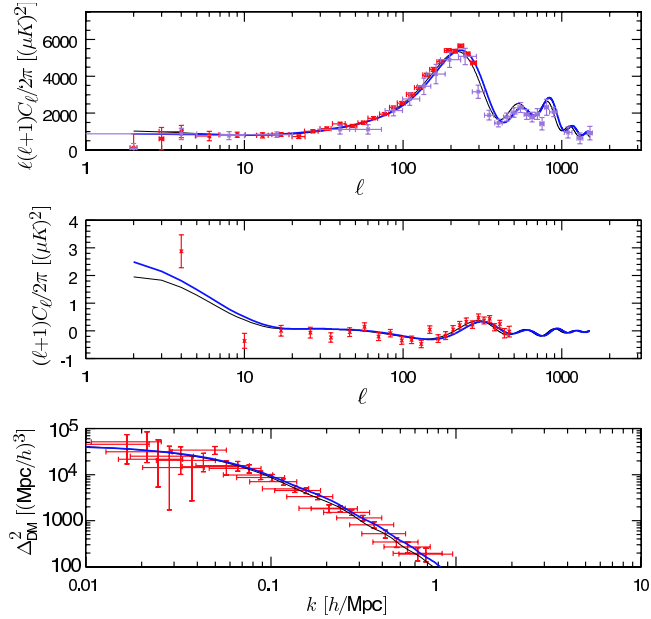}}
\caption{Angular power spectrum (top panel) and $T$-$E$ cross-correlation (middle panel) 
versus the
multipole $\ell$, and matter power spectrum versus the wave 
number $k$ (bottom panel), for
the periodic equation of state \eref{eq:oureos}  (blue line) and
for the $\Lambda$CDM model (black line).  The top panel shows
experimental data from the WMAP experiment
(red)~\cite{wmap} and  from the combination of all CMB data
(purple)~\cite{Wang:2002rt}.  The middle panel shows WMAP data. 
The data in the bottom panel are from an
independent analysis of the 2dF survey~\cite{Tegmark:2001jh}.}
\label{fig:cmbfast}
\end{figure}

To explore the implications of the periodic equation of state for
anisotropies of the cosmic microwave background, we have made the
appropriate modifications to the \textsc{cmbfast}~\cite{Seljak:1996is}
code, assuming three massless neutrinos and fixing the helium mass
fraction at $y_{\mathrm{He}} = 0.24$.  We show in \Fref{fig:cmbfast}
that the undulant universe describes the angular power spectrum,
temperature-polarization cross-correlation, and matter power spectrum
with the same degree of fidelity as the $\Lambda$CDM model. 
The structure at large scales ($\ell \ltap 10$) in the $T$-$E$ cross 
correlation is a consequence of 
reionization~\cite{deOliveira-Costa:2002rv}. Similar agreement can be 
expected for $b \ltap 2$: small values of the frequency parameter 
resemble the \lcdm\ picture, and values close to $b = 1$ yield similar 
averages of the vacuum-energy equation of state.

\subsection{SN Ia luminosity distance}
Type Ia supernovae now constitute an incisive probe of the state of 
the Universe near the present epoch. Existing data provide good 
resolution in the redshift range $0 \ltap z \ltap 1.7$. In the 
concordance (\lcdm) model inferred from these and other recent data, 
this is the range in which dark energy works its influence on the 
cosmic expansion rate.

The observational technique consists in determining the apparent 
magnitude $m$ (essentially the logarithm of the observed flux) and 
the redshift $z$. The apparent magnitude is related to the absolute 
magnitude $M$ of the supernova through the luminosity distance
\begin{equation}
    d_{\mathrm{L}} = c(1+z) \int_{0}^{z} 
    \frac{dz^{\prime}}{H(z^{\prime})} \; ,
    \label{eq:lumdist}
\end{equation}
as\footnote{The numerical factors are astronomical conventions.}
\begin{equation}
    \mu \equiv m - M = 5 \log_{10} \left( 
    \frac{d_{\mathrm{L}}}{1\mpc}\right) + 25 \;.
    \label{eq:apparentmag}
\end{equation}
The resolution hinges on establishing the absolute magnitude as a 
standard candle.

\begin{figure*}[tb]
\centerline{\includegraphics[width=9cm]{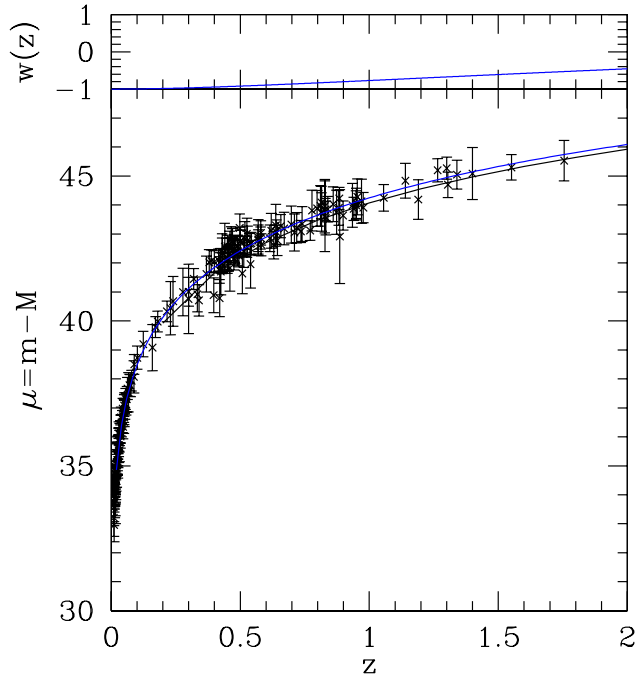}}
\caption{ The distance modulus $\mu = m - M$ for the \lcdm\ model 
(black line) and for the undulant universe of \Eref{eq:oureos} (blue 
line), compared with the luminosity modulus--redshift data from the SN 
gold and silver samples~\cite{Riess:2004nr}. The periodic equation of 
state is plotted in the upper panel.}
\label{fig:sn1}
\end{figure*}

\begin{figure*}[tb]
\centerline{\includegraphics[width=7.5cm]{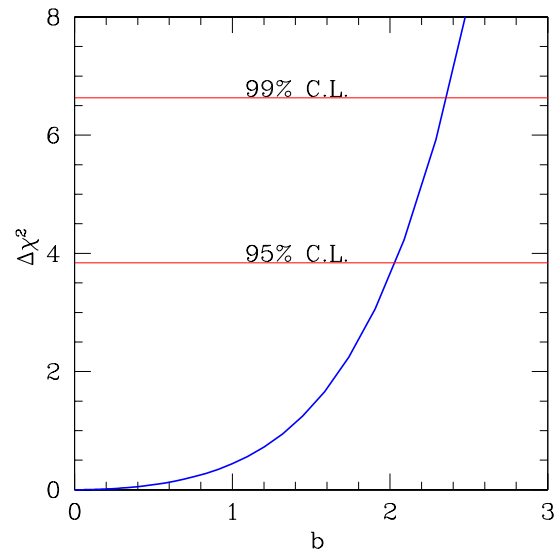}}
\caption{{ $\Delta \chi^2$ as a function of the $b$ parameter for 
one-parameter fits to the luminosity distance modulus.
Red lines indicate the bounds at $95\%$ and $99\%$ C.L.}}
\label{fig:sn2}
\end{figure*}

In  \Fref{fig:sn1}, we compare the luminosity distance
modulus of the Supernova Search Team's gold and silver samples with the
expectations of the \lcdm\ picture and the undulant universe
\eref{eq:oureos}.  For redshifts in the range $0 \ltap z \ltap 2$, the
models cannot now be distinguished.  To quantify the range of 
frequency parameters $b$ that adequately reproduce the existing data, 
we have modified Yun Wang's supernova flux-averaging likelihood 
code~\cite{wang1,wang2}  to incorporate the periodic equation of 
state \eref{eq:oureos2}. In \Fref{fig:sn2} we show how the
goodness of fit to the gold and silver sample data, measured 
by $\chi^{2}$, depends upon $b$. The best fit corresponds to $b = 0$, 
the \lcdm\ solution, but a frequency parameter as large as $b = 2$ is 
tolerated at 95\% C.L. In particular, our canonical undulant model 
\eref{eq:oureos} delivers a $\chi^{2}$ that differs only 
insubstantially from the minimum value.

\begin{figure*}[tb]
\centerline{\includegraphics[height=6cm]{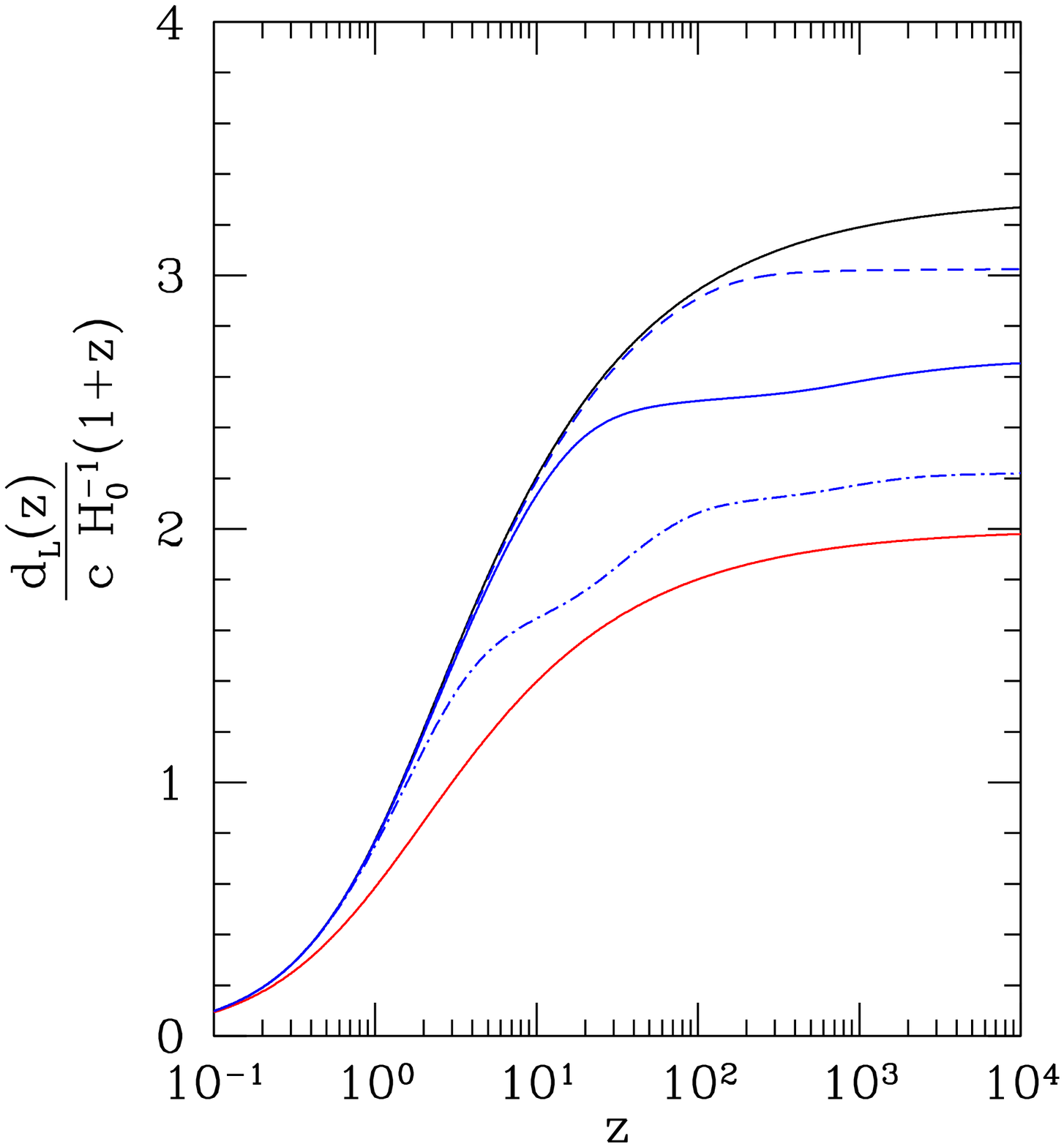}
 \includegraphics[height=6cm]{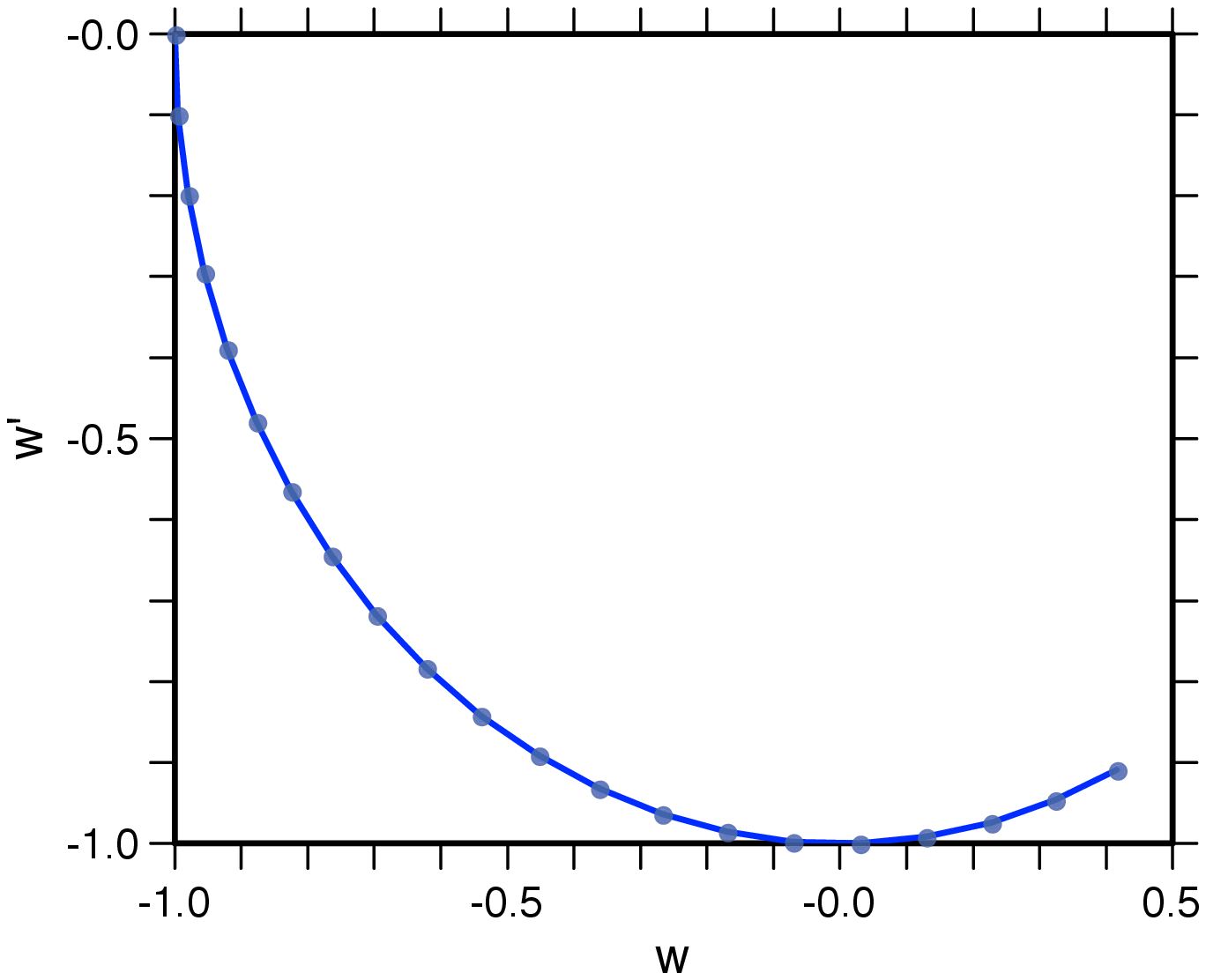}}
\caption{{The left panel displays the supernova luminosity distance \textit{versus}
redshift for the \lcdm\ (black), undulant universe of \Eref{eq:oureos}
(solid blue), critical SCDM universe (red).  Also shown are two variants of the undulant 
universe, defined by \eref{eq:oureos2} with $b = 2$ (dot-dashed blue) 
and $b = 0.6$ (dashed blue). The right panel shows the correlation
between the equation of state $w$ and its derivative $w^{\prime} = 
dw/d\ln{a}$, for the periodic equation of state \eref{eq:oureos2}. 
The point at $(-1,0)$ corresponds to zero redshift, $\ln{a}=0$, and 
the step between points is $\Delta\ln{a} = -0.1/b$.}}
\label{fig:sn3}
\end{figure*}

The equation of state \eref{eq:oureos}, plotted in the upper panel of
\Fref{fig:sn1}, does increase from $w(z=0) = -1$ to $w(z=2) = -0.45$,
so future high-precision measurements might make the distinction.  We
show the possibilities for experimental discrimination in two ways in
\Fref{fig:sn3}.  The left panel shows that the difference between the
implications of \lcdm\ and the undulant universe becomes more
pronounced with increasing redshift.  The canonical undulant universe
with $b= 1$ deviates appreciably from the \lcdm\ for redshifts $z \gtap
10$; the departure occurs in the neighborhood of $z = 3$ for $b=2$ and
at $z \sim \mathcal{O}(100)$ for $b = 0.6$.  As is well known, the
critical (SCDM) universe characterized by $\Omega_{m}=1$ is ruled out
by observations in the neighborhood of $z = 1$.  

The right panel of \Fref{fig:sn3} shows the correlation~\cite{Linder:2002et} 
between the equation of state, $w$, and its derivative, $w^{\prime} \equiv 
dw/d\ln{a}$ for the periodic equation of state \eref{eq:oureos2} near 
the present. The correlation is independent of the frequency 
parameter $b$, but the association of a particular value of $w$ with 
scale factor of course depends on the periodicity, as detailed in the 
figure caption.

Significant improvements are promised by candidates for the NASA/DOE 
Joint Dark Energy Mission~\cite{jdem}: DESTINY~\cite{destiny}, 
JEDI~\cite{jedi}, and SNAP~\cite{snap}, alone and in combination with 
cosmic microwave background results from the European Space Agency's Planck 
satellite~\cite{planck}.
\subsection{Galaxy clusters} 
Measurements of the apparent redshift dependence of the baryonic mass 
fraction of galaxy clusters can be used to constrain the geometry of 
the Universe and, hence, the amount and character of dark energy. The 
geometry enters in the dependence of the baryonic mass fraction on the 
assumed angular diameter distances, $d_{\mathrm{A}} = 
d_{\mathrm{L}}/(1+z)^2$, to the clusters. The baryonic mass fraction 
in the largest clusters should be independent of redshift, provided 
that the reference cosmology used in making the baryonic mass fraction 
measurements matches the true cosmology of the physical Universe.

\begin{figure*}[tb]
\vspace{1.0cm}
\centerline{\includegraphics[width=6.25cm]{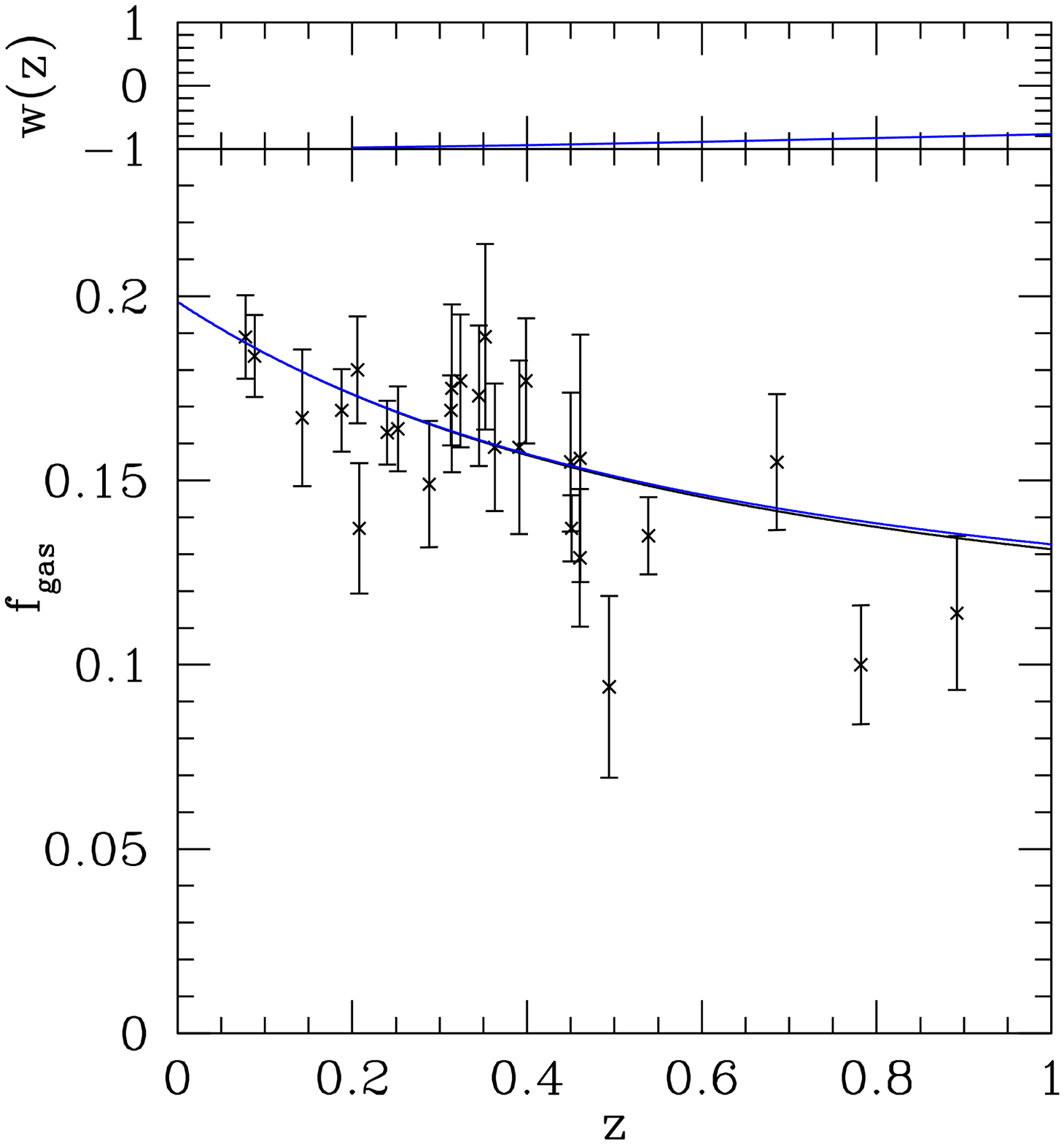}
\includegraphics[width=6.5cm]{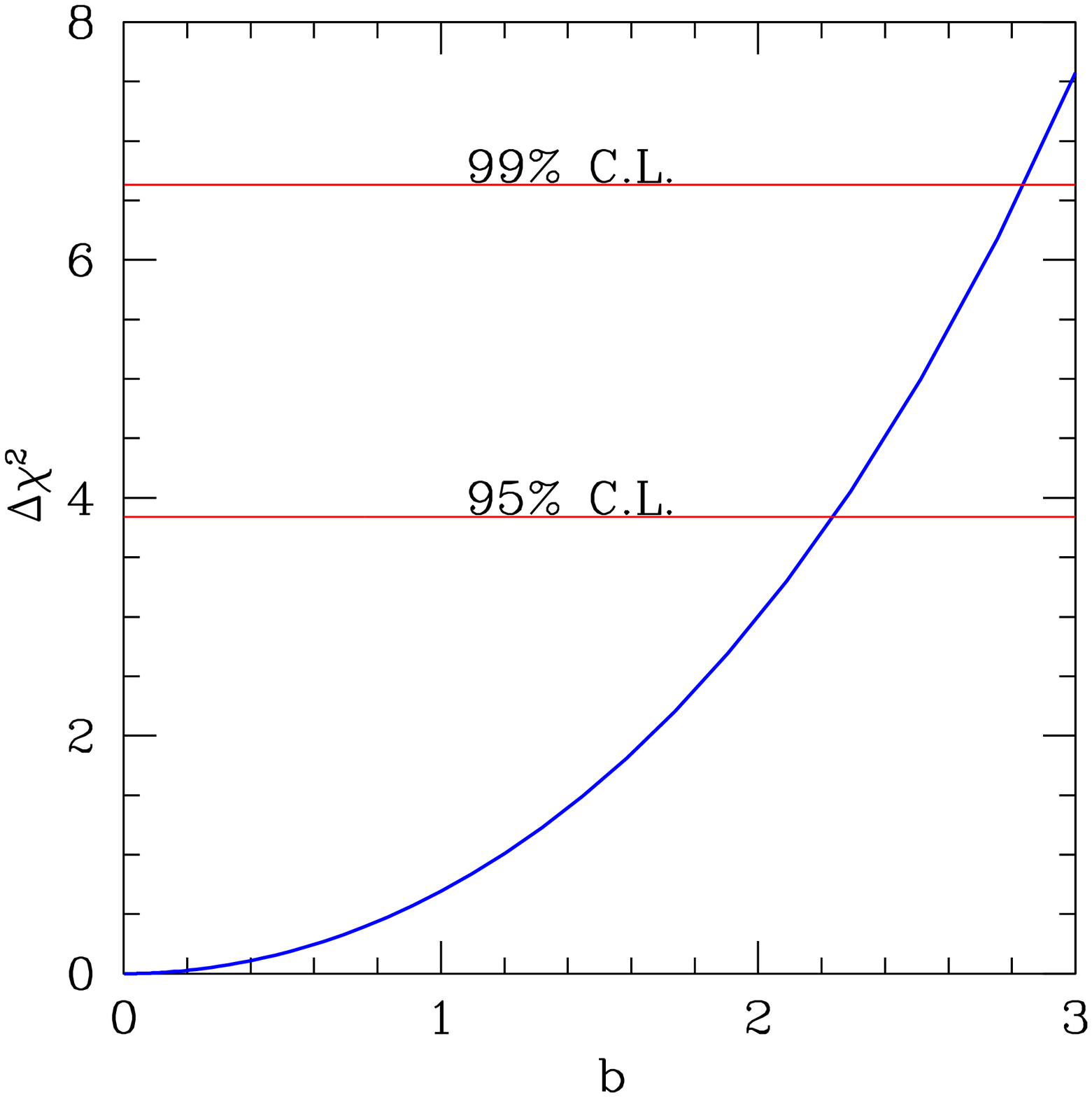}}
\caption{{(Left panel) Apparent variation of the X-ray gas mass
fraction (with $1\sigma$ rms errors) as function of the redshift for
the reference SCDM. The expection of the \lcdm\ cosmology is plotted as
the black curve.  The blue curve shows the prediction of the undulant
universe \eref{eq:oureos}.  The periodic equation of state is plotted 
in the upper panel. (Right panel) $\Delta \chi^2$ results for
the analysis of the X-ray mass fraction data as function of the
frequency parameter $b$ of \eref{eq:oureos2}.}} \label{fig:chandra1}
\end{figure*}

Because galaxy clusters are so large, it is plausible that they
represent a fair sample of the matter in the universe, so that the
relative amounts of hot gas and dark matter should be the same for
every cluster.  Moreover, the baryonic-to-total mass in the clusters
should closely match the ratio of the cosmological parameters
$\Omega_{b}/\Omega_{m}$.  By measuring the X-ray emissivity of 26
dynamically relaxed galaxy clusters in the redshift range $0.07< z
<0.9$ with the Chandra X-ray Observatory~\cite{chandra}, Allen and
collaborators have determined the X-ray gas mass fraction in these
systems, using the SCDM reference cosmology~\cite{Allen:2004cd}.  The
variation of the inferred gas fractions with redshift, shown in the
left panel of \Fref{fig:chandra1},
indicates that the distance-redshift correlation of the SCDM reference
cosmology does not correspond to our Universe.  

To compute the expectations of cosmologies that include vacuum energy, 
we evaluate the gas mass fraction,
\begin{equation}
f_{\rm gas}(z) = \frac{ \mathcal{B}\, \Omega_{\rm b}} {\left(1+0.19
\sqrt{h}\right) \Omega_{\rm m}} \left[ \frac{d_{\rm
A}^{\rm SCDM}(z)}{d_{\rm A}^{\mathrm{vac}}(z)} \right]^{1.5}\;.
\label{eq:fgas}
\end{equation}
The bias factor $\mathcal{B} = 0.824\pm 0.089$, for which we adopt a
Gaussian prior~\cite{Allen:2004cd,Rapetti:2004aa}, accounts for the
relatively small amount of baryonic material expelled from galaxy
clusters as they form.  The \lcdm\ model (black curve) and the periodic
equation of state \eref{eq:oureos} (blue curve) both reproduce the data
faithfully.  

The gas mass fraction is
sensitive not only to the amount of vacuum energy, but also to its
character, because the luminosity distance defined in \eref{eq:lumdist}
gives some sensitivity to the equation of state through $H(z)$. The right 
panel of \Fref{fig:chandra1} shows how the goodness of fit, as 
measured by $\chi^{2}$, depends on the frequency parameter $b$ of 
\Eref{eq:oureos2}. The current gas mass fraction measurements disfavor values 
$b \gtap 2.2$, so are slightly less restrictive constraint than the 
supernova observations.

\subsection{Structure formation \label{subsec:growth}} 
The development of large scale structure in the universe is sensitive
to the presence and character of dark energy.  Vacuum energy influences
the balance of attractive gravititational stability against the
dynamical friction of the expansion.  For an undulant universe, the
equation of state of the vacuum energy affects the cosmic volume in 
which structures form, and rules the dynamical behavior of the dark 
energy.

\begin{figure*}[tb]
\centerline{\includegraphics[width=10cm]{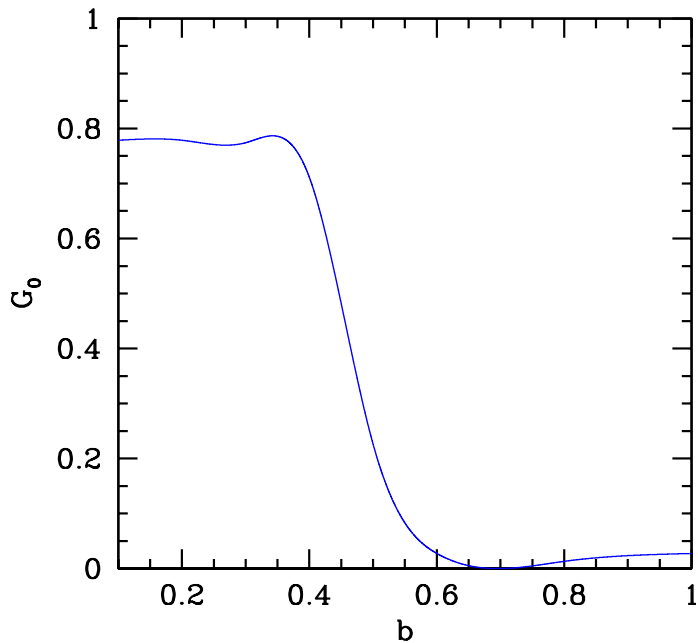}}
\vspace*{-12pt}
\caption{Linear growth factor in the present universe, 
$G_{0} \equiv G(a = 1)$ defined through \Eref{eq:normgrowth} for an 
undulant universe characterized by the equation of state 
\eref{eq:oureos2}, as a function of the frequency parameter $b$.} \label{fig:Gofb}
\end{figure*}

A simple, and useful, test is provided by the linear growth factor
considered by Linder and Jenkins~\cite{Linder:2003dr}.  Considering
linear evolution of the perturbations only, we may define the
normalized growth factor
\begin{equation}
G^{\prime\prime}(a)+\frac{3}{2}\left[\frac{7}{3}-
{\displaystyle\frac{w(a)}{1+X(a)}}\right]{G^{\prime}(a)\over a} 
+{3\over2}\cdot{1-w(a)\over 1+X(a)}\,{G(a)\over a^2}=0\,,
\label{eq:normgrowth}
\end{equation} 
where a prime denotes a derivative with respect to the scale factor 
$a$ and 
\begin{equation}
    X(a) = \frac{\Omega_{m}}{1 - \Omega_{m}} \cdot \frac{1}{g(a)}\,,
    \label{eq:Xdef}
\end{equation}
with $g(a)$ given by \eref{eq:gofa}, is the ratio of matter density 
to dark energy density when radiation is negligible. Then, 
adopting the boundary condition $G(0) = 1$, we may solve for 
the normalized growth factor $G(a)$. The resulting values at the 
current epoch ($a = 1$) are shown in \Fref{fig:Gofb} for the undulant 
universe with over the range of frequency parameters $0 \le b \le 1$.

\begin{figure*}[tb]
\centerline{\includegraphics[width=10cm]{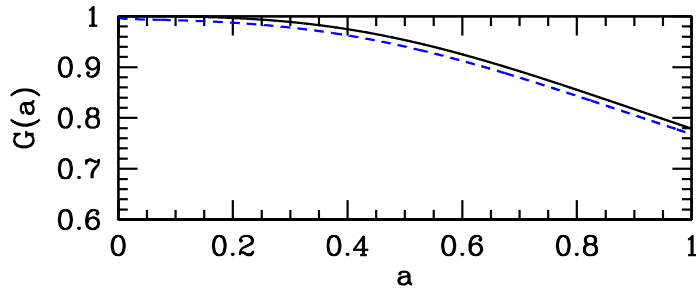}}
\vspace*{-12pt}
\caption{Dependence on scale factor $a$ of the linear growth factor
$G(a)$ for the \lcdm\ cosmology (solid line) and for 
 an undulant universe with frequency parameter $b = 0.4$ (dashed line).} 
\label{fig:bigGofa}
\end{figure*}

In the limit as $b \to 0$, the periodic equation of state
\eref{eq:oureos2} approaches the \lcdm\ cosmology, so reproduces the
canonical growth factor.  The value of the growth factor in the present
universe depends little on the frequency parameter for $b \ltap 0.4$,
then drops precipitously to small values.  In particular, the case $b =
0.4$, the consequences of which are in excellent agreement with all the
other constraints we have examined, hardly differs from the \lcdm\
result.  Current determinations of $\sigma_{8}$ agree with the \lcdm\ expectation
within the uncertainties of about 20\%. Indeed, the $b \ltap 0.4$ undulant 
universes track the \lcdm\ solution as a function of the scale 
factor $a$. We display the $b = 0.4$ and \lcdm\ solutions in 
\Fref{fig:bigGofa}.

The undulant universe characterized by frequency parameter $b = 1$
fails the test of the linear growth factor, as commented by
Linder~\cite{Linder:2005in}.  While taking note of this disagreement,
we believe it is prudent to note that the strategy of considering
linear evolution of the perturbations, while plausible, has not been
exhaustively validated.  The undulant universe
of \Eref{eq:oureos} does reproduce the matter power spectrum determined
by cosmic microwave background measurements, so it is possible that a
different treatment of structure formation might have a favorable
outcome.  Alternatively, it may be that frequency parameters $b \gtap
0.4$ can be brought into agreement with structure formation constraints
only by delaying the emergence of the vacuum energy component or, more
generally, by introducing coupling between the vacuum energy and other
components, as in the slinky inflation model.

\section{Outlook}
The undulant universe offers a new response to the cosmic coincidence
problem: the current state of the Universe, with $\Omega_{m} \approx
\Omega_{v}$ and $w_{v} \approx -1$, has happened before and will happen
again, so long as the frequency parameter $b \gtap 0.6$ in the undulant
equation of state (\ref{eq:oureos2}).  No fine tuning is required, in
the sense that $0.5 \le \Omega_{v} \le 0.9$ with $w_{v} \le -0.7$
occurs with $\sim 9\%$ probability for $10^{-7} \le a \le 10^{+7}$ for
the prototype undulant universe with $b=1$. The corresponding number 
is 4.5\% for the case $b=0.4$. [See the details in
Table~\ref{tbl:stats}.]

We find that periodic equations of state \eref{eq:oureos2} reproduce 
the correct power spectrum and CMB temperature asymmetries, so long 
as the frequency parameter $b \ltap 2$. Moreover, such models satisfy 
constraints on the matter density at the time of big bang 
nucleosynthesis, except for frequency parameters in the neighborhood 
of $b \approx \cfrac{1}{4}$. The undulant cosmologies fit very well 
late-time geometry probes such as supernova luminosity distances; at 
95\% C.L., frequency parameters as large as $b = 2$ are acceptable. 
The one delicate issue is structure formation as embodied in the 
growth factor. A linear treatment of the evolution of density 
perturbations restricts the frequency parameter to $b \ltap 0.4$. The 
case of $b = 0.4$ is in comfortable agreement with all observations, 
but entails only the present matter--dark-energy coincidence since the 
radiation-dominated era ended.

We have postulated the periodic equation of state, not derived it from 
a dynamical principle, in the interest of exploring alternative 
implications of the discovery of the accelerating universe. The more 
ambitious program of slinky 
inflation~\cite{Barenboim:2005np} gives an explicit construction of a 
periodic equation of state from a potential that governs the behavior 
of a scalar field~\cite{Nakamura:1998mt,Gerke:2002sx,Guo:2005at}. 

On the observational front, it is of clear interest to pin down the
vacuum-energy equation of state $w(z)$ and to seek evidence that it
varies with time, for redshifts $z \approx 1$~\cite{Alam:2003fg,Jassal:2005qc}.  
This observational challenge is an imperative for quintessence models 
in general, for the \lcdm\ picture, and for the undulant universe and 
its extension, slinky inflation. The observational survival of 
periodic equations of state highlights the fact that, at present, we 
have telling observations at only a few epochs in the history of the 
universe. It is of clear interest to devise new observational tests 
that will explore new ranges in redshift, making new strata of the 
fossil record available for our scrutiny. The wide excursions in the 
deceleration parameter at different epochs, shown in Figure~\ref{fig:decel}, 
offer encouragement for probing deeper in redshift than is possible 
with supernovae.

While no finite set of astronomical measurements made over a finite
time will ever allow us to determine the ultimate fate of our
Universe~\cite{Krauss:1999br}, we can hope to look some distance into
the future.  Wang \& collaborators have quantified~\cite{Wang:2004nm}
the limited reach of reliable extrapolations in the framework of the
simplest doomsday model, in which the universe collapses swiftly, once
it ceases to expand.  They reckon the collapse time
$t_{\mathrm{collapse}} \gtap 42 (24)\gyr$ from today at 68\% (95\%)
C.L. An interesting parameter in the undulant universe is the moment
that marks the onset of the next period of deceleration, which we
characterize as the moment at which the deceleration parameter next
exceeds the value that obtains in a matter-critical universe,
$q_{\mathrm{SCDM}} = \cfrac{1}{2}$.  That circumstance occurs at $(11,
23, 39, 153)\gyr$ in the future for frequency parameters $b = (3, 2,
1, 0.4)$. The cases with $b \ltap 2$ all respect the doomsday bounds 
of Ref.~\cite{Wang:2004nm}.

The undulant universe explored here serves as a reminder that the range
of possible destinies for the Universe, even in the near term, is very
broad indeed.  The
universe need not necessarily evolve toward the cataclysm of terminal
inflation or recollapse, but might steer a middle course not so
different, on average, from a critical universe dominated by matter.
The main lesson of the undulant universe is that it is premature to
anoint the \lcdm\ model as the sole candidate for the new standard
cosmology.

\ack 
Fermilab is operated by Universities Research Association Inc.\ under
Contract No.\ DE-AC02-76CH03000 with the U.S.\ Department of Energy.
It is a pleasure to thank Steve Allen, Eric Linder, Joe Lykken and Matias
Zaldarriaga for helpful communications.

\end{document}